\documentclass[aps,prb,superscriptaddress,twocolumn,showpacs,amsmath,amssymb]{revtex4}
\usepackage{stmaryrd}
\usepackage{times}
\usepackage{graphicx}
\usepackage{dcolumn}
\usepackage{bm}
\usepackage{epstopdf}
\usepackage[none]{hyphenat}

\begin{document}


\title{Band structure and spin texture of Bi$_2$Se$_3$/3d ferromagnetic metal interface}

\author{Jia Zhang}
\email[Email:]{jzhang48@unl.edu}
\affiliation{Department of Physics and Astronomy, University of Nebraska, Lincoln, Nebraska 68588, USA}

\author{Julian P. Velev}
\affiliation{Department of Physics and Astronomy, University of Puerto Rico, San Juan, Puerto Rico 00931, USA}
\affiliation{Department of Physics and Astronomy, University of Nebraska, Lincoln, Nebraska 68588, USA}

\author{Xiaoqian Dang}
\affiliation{Department of Physics and Astronomy, University of Nebraska, Lincoln, Nebraska 68588, USA}

\author{Evgeny Y. Tsymbal}
\email[Email:]{tsymbal@unl.edu}
\affiliation{Department of Physics and Astronomy, University of Nebraska, Lincoln, Nebraska 68588, USA}

\date{\today}

\begin{abstract}

\noindent
The spin-helical surface states in three-dimensional topological insulator (TI), such as Bi$_2$Se$_3$, are predicted to have superior efficiency in converting charge current into spin polarization. This property is said to be responsible for the giant spin-orbit torques observed in ferromagnetic metal/TI structures. In this work, using first-principles and model tight-binding calculations, we investigate the interface between the topological insulator Bi$_2$Se$_3$ and 3d-transition ferromagnetic metals Ni and Co. We find that the difference in the work functions of the topological insulator and the ferromagnetic metals shift the topological surface states down about 0.5 eV below the Fermi energy where the hybridization of these surface states with the metal bands destroys their helical spin structure. The band alignment of Bi$_2$Se$_3$ and Ni (Co) places the Fermi energy far in the conduction band of  bulk Bi$_2$Se$_3$, where the spin of the carriers is aligned with the magnetization in the metal. Our results indicate that the topological surface states are unlikely to be responsible for the huge spin-orbit torque effect observed experimentally in these systems.

\end{abstract}


\maketitle


\noindent
Topological insulators (TIs) represent a state of matter which is insulating in the bulk, but exhibits gapless surface states protected by time reversal symmetry \cite{Moore:2010, Hasan:2010, Qi:2011}. Three-dimensional (3D) TIs have been discovered among the Bi$_2$Se$_3$ family of materials. Their properties were elaborated from first-principles calculations \cite{Murakami:2007, Zhang:2009, Zhang:2010, Yazyev:2010} and the signature Dirac cone was observed experimentally in the angle-resolved photoemission spectroscopy (APRES) experiments \cite{Hsieh:2008, Hsieh:2009, Wray:2010}. A key feature of carriers in the topologically protected surface state is the locking between their spin and momentum, which are constrained to be perpendicular by the strong spin-orbit coupling (SOC)  \cite{Hasan:2010, Qi:2011}. This property has been proposed as the foundation of applications based on TIs.

Magnetization switching in nanoscale devices induced by the electric current is one of the most researched applications in recent years \cite{Ohno:2000}. In particular, the magnetization switching via spin-orbit torque (SOT) has been demonstrated in ferromagnetic (FM) layers interfaced with heavy metals with strong SOC, such as Pt \cite{Miron:2010, Miron:2011} or Ta \cite{Buhrman:2012}. There are two physical mechanisms contributing to SOT. First, when the charge current passes along the FM layer, the SOT is produced by the interfacial Rashba SOC \cite{Miron:2010}. Second, when the current passes along the adjacent metal layer, the bulk spin Hall effect (SHE) in that layer produces spin accumulation at the interface exerting SOT on the magnetization of the ferromagnet \cite{Buhrman:2012}.

It has been proposed that the spin-helical surface states of TIs would be more efficient than heavy metals with strong SOC in converting charge to spin current and producing SOT \cite{Fisher:2016}. Indeed, there have been reports of giant SOT arising at the interface between a topological insulator and ferromagnet (NiFe or CoFeB) \cite{Mellnik:2014, Wang:2015} or another magnetically-doped topological insulator (Cr-doped BiSbTe$_3$) \cite{Fan:2014}. The explanation for this high efficiency is rooted in the assumption that the charge current is carried predominantly by the surface state of the TI and the spin accumulation produced by the spin-momentum locking is much larger than it would be in non-topological materials with SOC of a similar magnitude \cite{Mellnik:2014, DasSarma:2010, Burkov:2010}.

Due to the importance of the spin texture of the surface state, the question arises of its robustness when TI is interfaced with a metal. The spin texture of the pure surface state in Bi$_2$Se$_3$ has been studied in model and first-principles calculations \cite{Zhang:2010, Yazyev:2010, Zhang2:2013, Guo:2011}. It has been shown theoretically that the surface state is robust with respect to disorder, if the perturbation is sufficiently small \cite{Guo:2012}, as well as with respect to moderate doping with Mn atoms \cite{Mertig:2012}. Similarly, it has been demonstrated from model calculations that at TI/metal interfaces the surface state survives, provided that the bonding is weak \cite{DasSarma:2010}.  The surface transport in Bi$_2$Se$_3$ has been confirmed experimentally by measuring the thickness dependence of the conductance of TI slabs \cite{Oh:2012}. Spin-polarization has been probed by injecting energetic electrons in Bi$_2$Se$_3$ \cite{Liu:2015}. However, band gap opening and magnetization reorientation associated with Mn doping have been reported \cite{Xu:2012}.


Thus, the existing studies of the robustness of the spin-helical surface states are mostly limited, so far, to dilute doping with impurities and weak coupling regime. The properties of TI/FM metal interfaces have not been studied in sufficient detail. In this paper, we use density-functional calculations to investigate the electronic and magnetic structure of Bi$_2$Se$_3$ interfaces with the FM metals Ni and Co. In order to elucidate the mechanism of the interaction of the TI surface state with the FM metal, we complement the first-principles calculations with tight-binding model studies of TI/FM metal and insulator interfaces. Our results show that the strong hybridization at the TI/FM metal interface largely destroys the topologically protected state with its spin-helical structure.

\begin{figure}[h]
\includegraphics[width=8.6cm]{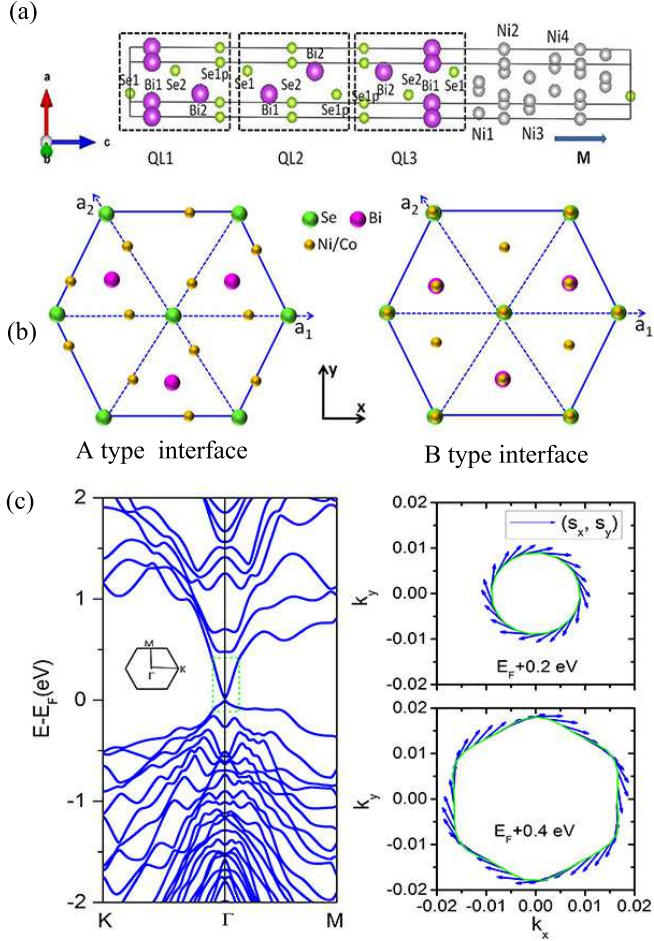}
\caption{(a) Supercell structure of Bi$_2$Se$_3$(3QLs)/Ni,Co(6MLs) (b) Alternative interface terminations at Bi$_2$Se$_3$ (0001) and Ni, Co (111) interface. (c) Band dispersion of 3QL-Bi$_2$Se$_3$ (0001) slab along the high symmetry lines (left). Spin texture of TI surface states at $E=E_{F}+0.2$ eV and $E=E_{F}+0.4$ eV (right).}  \label{fig1}
\end{figure}

\noindent
\emph{First-principles calculations}:
The atomic structure model for the Bi$_2$Se$_3$/Ni(111) interface is shown in Fig.~\ref{fig1}(a). The Bi$_2$Se$_3$ has hexagonal symmetry and the atomic structure along the (0001) direction consists of quintuple layers (QLs) of two Bi layers sandwiched between three Se layers. We use the experimental lattice constants $a=4.138$ {\AA} and $c=28.640$ {\AA}  \cite{Zhang:2010}. The $\sqrt{3} \times \sqrt{3}$ rotated (111) surface of the fcc Ni matches the Bi$_2$Se$_3$ with less than 4.1\% mismatch. The two interface terminations considered are shown in Fig.~\ref{fig1}(b). The same structural model is used for constructing the Bi$_2$Se$_3$/Co(111) interface.

The calculations are carried out by using the Vienna \emph{ab initio} simulation package (VASP) \cite{Kresse:1996}. The Perdew-Burke-Ernzerhof (PBE) generalized gradient approximation is used for the exchange-correlation potential \cite{PBE:1996}. The calculations are performed in the presence of SOC \cite{VASP:soc1, VASP:soc2}. The self-consistency cycle is run using $16 \times 16 \times 1$ k-point grid in the Brillouin zone and the plane-wave energy cutoff of 450 eV. The density of states (DOS) and the magnetocrystalline anisotropy energy (MAE) are calculated with a finer $32 \times 32 \times 1$ k-point grid. Atomic coordinates and the supercell dimension along the c-axis direction are allowed to relax until the force on all atoms is less than 1 meV/{\AA}.

First, we calculate the band structure of a free-standing 3QL-Bi$_2$Se$_3$ (0001) slab. The results are displayed in Fig.~\ref{fig1}c. There are two surface states localized at top and bottom surfaces, which have the opposite spin helicities \cite{Yazyev:2010}. The spin of each Bloch state is evaluated through the expectation value of the spin operator $S_{n,\mu}(\vec{k}) = \frac{\hbar}{2} <{\psi_n}(\vec{k})| \sigma_{\mu} |{\psi_n}(\vec{k})> ({\mu}=x,y,z)$. The calculated spin textures corresponding to the top surface states of the slab at 0.2 and 0.4 eV above the Fermi energy are shown in Fig.~\ref{fig1}c (right panels). The spins lie entirely in the plane with the out-of-plane component $S_z$ being negligible. The important consequence of such a spin texture is that it carries spin density when charge current flows on the surface, resulting in the transport spin polarization \cite{Mellnik:2014}.

Next, consider the Bi$_2$Se$_3$/Ni(Co) supercell which contains 3 QLs of Bi$_2$Se$_3$ and 6 monolayers (MLs) of Ni or Co (Fig.~\ref{fig1}a). The A-type interface is found to be more energetically favorable (Fig.~\ref{fig1}b) with a lower total energy by 0.5 eV. In Table ~\ref{tab1} we summarize the properties of the interface, and to put them in perspective, compare them to the well-studied graphene/Ni(Co) \cite{Giovannetti:2008} and Fe/MgO \cite{JD:2006} interfaces. The calculated equilibrium interface distances between Se and Ni (Co) planes is 2.05 (2.10) {\AA}. The work of separation $W_{sep}$ for Bi$_2$Se$_3$/Ni(Co) is larger than that for graphene/Ni(Co) and comparable to that for the CoFe/MgO interface, which indicates strong bonding at the Bi$_2$Se$_3$/Ni(Co) interface. The MAE per interface is calculated as ${\rm MAE} = \frac{1}{2} (E_{M_{x}} - E_{M_{z}})$, where $E_{M_{x}}$ and $E_{M_{z}}$ are the total energies for magnetization pointing along the $x$ and $z$ directions, respectively. Within this definition, positive MAE implies the easy axis pointing along $z$ direction (i.e. perpendicular anisotropy). As is evident from Table ~\ref{tab1}, the Bi$_2$Se$_3$/Ni(111) interface exhibits in-plane anisotropy, whereas the Bi$_2$Se$_3$/Co(111) interface exhibits out-of-plane (perpendicular) anisotropy.


\begin{table}
\caption{Calculated interface properties: the equilibrium interface layer distance $d_0$, work of separation $W_{sep}$, and MAE. The corresponding reference values
for (Ni,Co)/Gr(graphene) and Fe/MgO interface are listed for comparison.} \label{tab1}
\centering
\begin{ruledtabular}
\begin{tabular}{c|c|c|c|c|c}
  Interface&Bi$_2$Se$_3$/Ni & Bi$_2$Se$_3$/Co & Gr/Ni & Gr/Co&Fe/MgO\\
  \hline
  mismatch&4.1\%&4.7\%&1.3\%&1.9\%&-3.8\%\\
  \hline
  $d_0$({\AA})&2.05&2.11&2.05&2.05&2.167\\
  \hline
  $W_{sep}$(J/m$^2$)&1.10&1.08&0.386&0.495&0.97-1.36\\
  \hline
  MAE(erg/cm$^2$)&-2.18&0.52&-&-&1-1.5
\end{tabular}
\end{ruledtabular}
\end{table}

\begin{figure*}
\includegraphics[width=16cm]{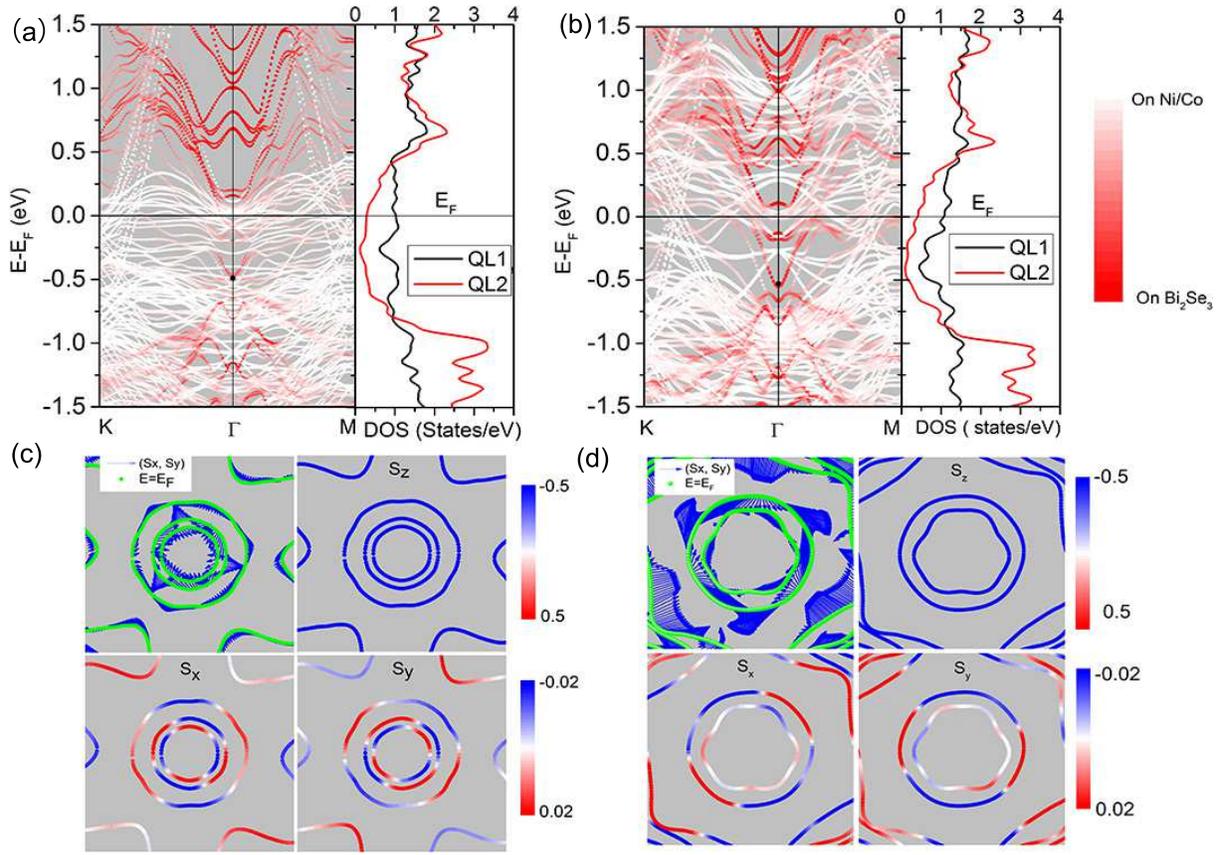}
\caption{Band structure and corresponding density of states (DOS) of the Bi$_2$Se$_3$/Ni(111) (a) and Bi$_2$Se$_3$/Co (111) (b) interfaces. The color intensity is proportional to the projected weight of the bands on Bi$_2$Se$_3$ (red) and Ni(Co) (white). The black dots indicate the position of the Dirac cone in a stand-alone Bi$_2$Se$_3$ (0001) surface. (c,d) The total spin (indicated by vectors) and spin components (S$_x$,S$_y$,S$_z$) indicated in color at $E=E_F$ around the Brillouin zone center for Bi$_2$Se$_3$/Ni(111) and Bi$_2$Se$_3$/Co(111), respectively.} \label{fig2}
\end{figure*}

Figs.~\ref{fig2}a,b show the band structure of the Bi$_2$Se$_3$/Ni(111) and Bi$_2$Se$_3$/Co(111) supercells, respectively. Here the bands are projected onto Bi$_2$Se$_3$ and Ni (Co), as indicated by the color. Due to the large electrostatic mismatch, charge flows from the metal into the conduction band of Bi$_2$Se$_3$, as seen from the relatively large DOS on the first QL of the TI (Figs.~\ref{fig2}a,b, right panels). Comparing the Bi$_2$Se$_3$ bands in the supercells (Figs.~\ref{fig2}a,b, red color) with those of the free-standing slab (Fig.~\ref{fig1}c) we observe that the remnant of the Dirac cone appears about 0.5 eV below the Fermi level. The position of the Dirac cone can be explained by the mismatch of the work functions of Bi$_2$Se$_3$(0001) and Ni(Co)(111). As seen from Table \ref{tab2}, the calculated work function of Bi$_2$Se$_3$ is 0.48 (0.51) eV higher than that of Ni (Co). Thus, the Dirac point falls fairly deep in the conduction band of the metals. This circumstance explains the fairly modest MAE calculated at these interfaces (Table \ref{tab1}). The MAE is expected to be most pronounced when the Dirac cone is around the Fermi level.

Further, we study the effect of the hybridization on the spin texture. Here and below we consider the magnetization of the ferromagnet (Ni or Co) pointing perpendicular to the plane, i.e. along the z direction. First, we look at the spin texture around the Fermi energy. The S$_x$, S$_y$, and S$_z$ projections for the spin of the bands at the Fermi energy near the Brillouin zone center are shown in Figs.~\ref{fig2}c,d for Ni and Co, respectively. We see that the spin is pointing along the magnetization direction of the ferromagnet and the in-plane spin components $S_{\|}$ show a complicated spin texture with absolute values one order of magnitude smaller than S$_z$. This result is not surprising due to absence of the Dirac states at the Fermi energy. The electrostatic mismatch at the interface places the Fermi energy far in the conduction band of the bulk Bi$_2$Se$_3$ where the hybridization between the TI and FM states induces an exchange splitting of the TI spin bands with an effective exchange field pointing along the FM magnetization. Transport as well as SOT properties are largely controlled by the Fermi surface, and therefore this result indicates that it is unlikely that the topological states contribute to these properties in the Bi$_2$Se$_3$/Co(Ni) (111) system.

\begin{figure*}
\includegraphics[width=16cm]{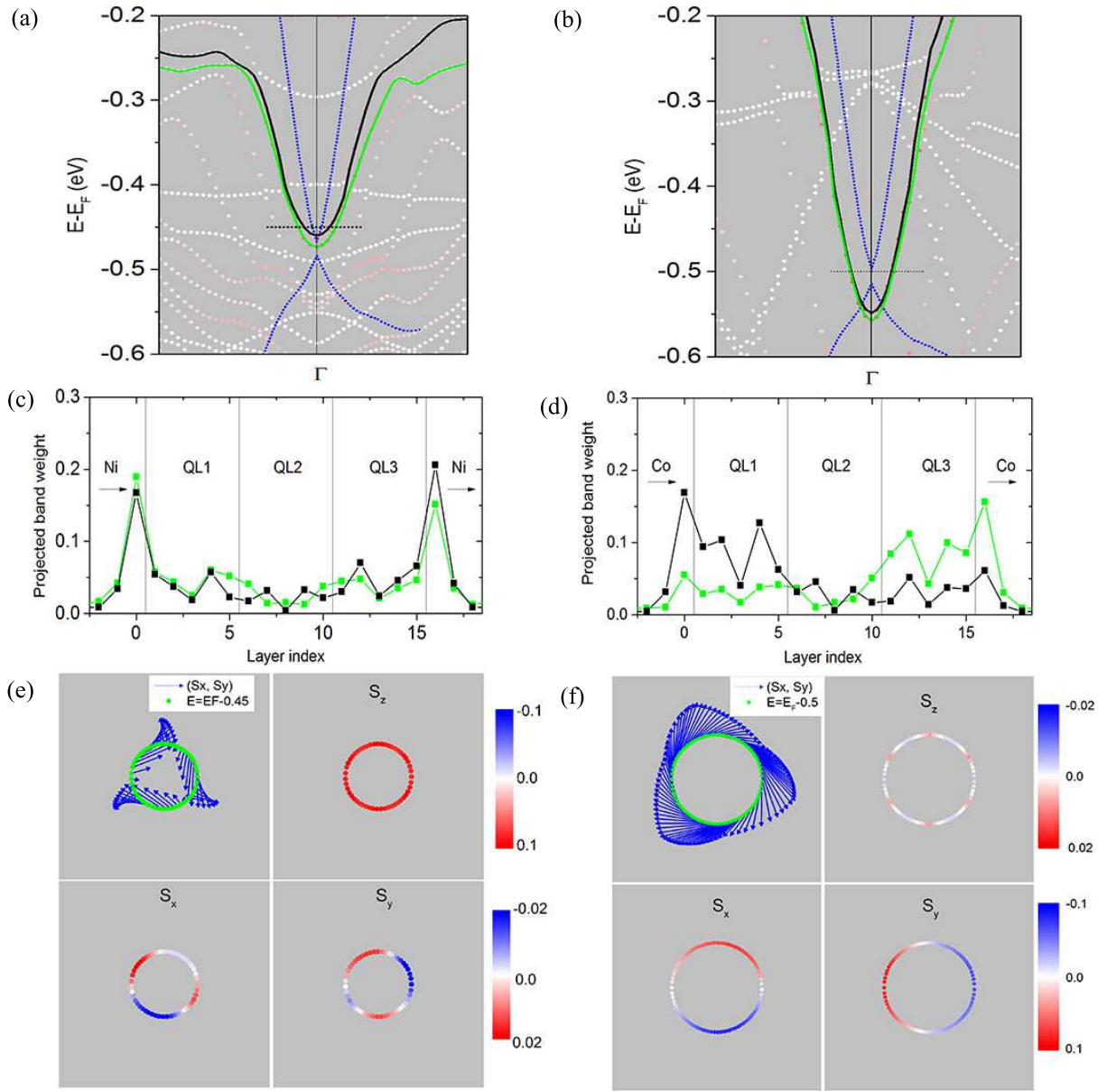}
\caption{Band structure of the Bi$_2$Se$_3$/Ni(111) (a) and Bi$_2$Se$_3$/Co(111)(b) interfaces around the remnant Dirac state. The color intensity is proportional to the projected weight of the bands on Bi$_2$Se$_3$ (red) and Ni(Co) (white). The black and green solid lines indicate the remnant of Dirac state at Bi$_2$Se$_3$/Co(Ni) interface. The dashed blue line indicates the TI surface state shifted according to the calculated work function mismatch. (c,d) The layer resolved band weight of the interface at the $\Gamma$ point. The arrows indicate the magnetic moment direction in ferromagnetic metals Ni and Co. (e,f) The spin texture of remnant Dirac states at the Bi$_2$Se$_3$/Ni(111) and Bi$_2$Se$_3$/Co(111) interfaces  at $E=E_{F}-0.45$ eV and $E=E_{F}-0.5$ eV as indicated by the horizontal dash black line in (a) and (b), respectively.} \label{fig3}
\end{figure*}

Now we discuss the remnant of the Dirac states lying at about 0.5 eV below the Fermi energy. As seen from Figs.~\ref{fig3}a,b, only the upper branch of the original topological state remains. The lower branch is smeared out both for Bi$_2$Se$_3$/Ni and Bi$_2$Se$_3$/Co interfaces. The upper branch consists of two subbands (indicated by black and green solid lines in Figs.~\ref{fig3}a,b), originating from the two interfaces in the supercell structure. The degeneracy of these subbands is lifted due to the effective exchange field produced by the magnetization of the ferromagnet pointing in the opposite direction with respect to the normal to each TI surface. The Dirac states at the Bi$_2$Se$_3$/Ni interface are hybridized stronger with the metal bands (due to their overlap with the Dirac states), as compared to the Bi$_2$Se$_3$/Co interface where the hybridization is weaker. By looking at the layer resolved band weight of these states at the $\Gamma$ point (Figs.~\ref{fig3}c,d), we see that the bands have higher weight at the interfaces, as expected for surface (interface) states. In case of Bi$_2$Se$_3$/Ni, however, each of the two bands has nearly equal weight at the two interfaces (Fig.~\ref{fig3}c), whereas in case of Bi$_2$Se$_3$/Co each of the two bands is peaked at one of the two interfaces (Fig.~\ref{fig3}d). The latter behavior is expected for the topologically protected surface states, whereas the former indicates that the two surface states are mixed through hybridization with the metal states.

Based on this result we can expect that the corresponding spin texture of the remnant Dirac bands in Bi$_2$Se$_3$/Ni and Bi$_2$Se$_3$/Co will be different. Indeed, as is evident from Fig.~\ref{fig3}(e) the spin texture of the Dirac state in Bi$_2$Se$_3$/Ni is completely destroyed, so that it has negligible in-plane spin components $S_x$ and $S_y$, whereas the $S_z$ component is large. For Bi$_2$Se$_3$/Co, however, as seen from Fig.~\ref{fig3}f, the spin texture of the Dirac state is largely preserved and the interface state has a helical spin structure with negligible $S_z$ component. We note however that in either case, due to of the remanent Dirac states lying far below the Fermi energy, they are not expected to contribute to the spin-dependent transport and SOT.

\begin{table}
\caption{Work functions of Bi$_2$Se$_3$ (0001), Co (111) and Ni (111) slabs. The atomic structure of the Ni and Co slabs is taken from the relaxed Bi$_2$Se$_3$/Ni(Co) supercell structure.} \label{tab2}
	\centering
	\begin{ruledtabular}
		\begin{tabular}{c|c|c|c|c}
			Material&3QL-Bi$_2$Se$_3$&Ni slab&Co slab&Gr\\
			\hline
			W(eV)  GGA+soc (This work) &5.67&5.19&5.16&-\\
			\hline
			Wref(eV) LDA+nsoc&-&5.47&5.44&4.48\\
			\hline
			W$_{exp.}$(eV)&-&5.35&5.0&4.6
		\end{tabular}
	\end{ruledtabular}
\end{table}

In order to see how the topological surface state recovers when the interface coupling strength is decreasing, we have calculated the band structure of Bi$_2$Se$_3$/Ni(111) while gradually increasing the interface distance between TI and the ferromagnet with respect to equilibrium by $\delta d$ (Fig. 1a, Supplemental Material). We find that with the increasing separation the hybridization decreases and respectively the Dirac cone becomes more pronounced and moves up in energy. When the interface separation is increased to be ${\delta}d=3.0$ {\AA}, the dispersions expected for the topological surface states appear  with a 30 meV gap due to the magnetic proximity effect. However, the helical spin texture is still not recovered (Figs. 1b,c, Supplementary Material). Thus, even a very weak interface coupling seems to be detrimental to the spin texture of the surface state of the TI. Only at interface separations of as large as ${\delta}d=8$ {\AA}, the surface state resembles that of a free slab.

\begin{figure}[h]
\includegraphics[width=8.6cm]{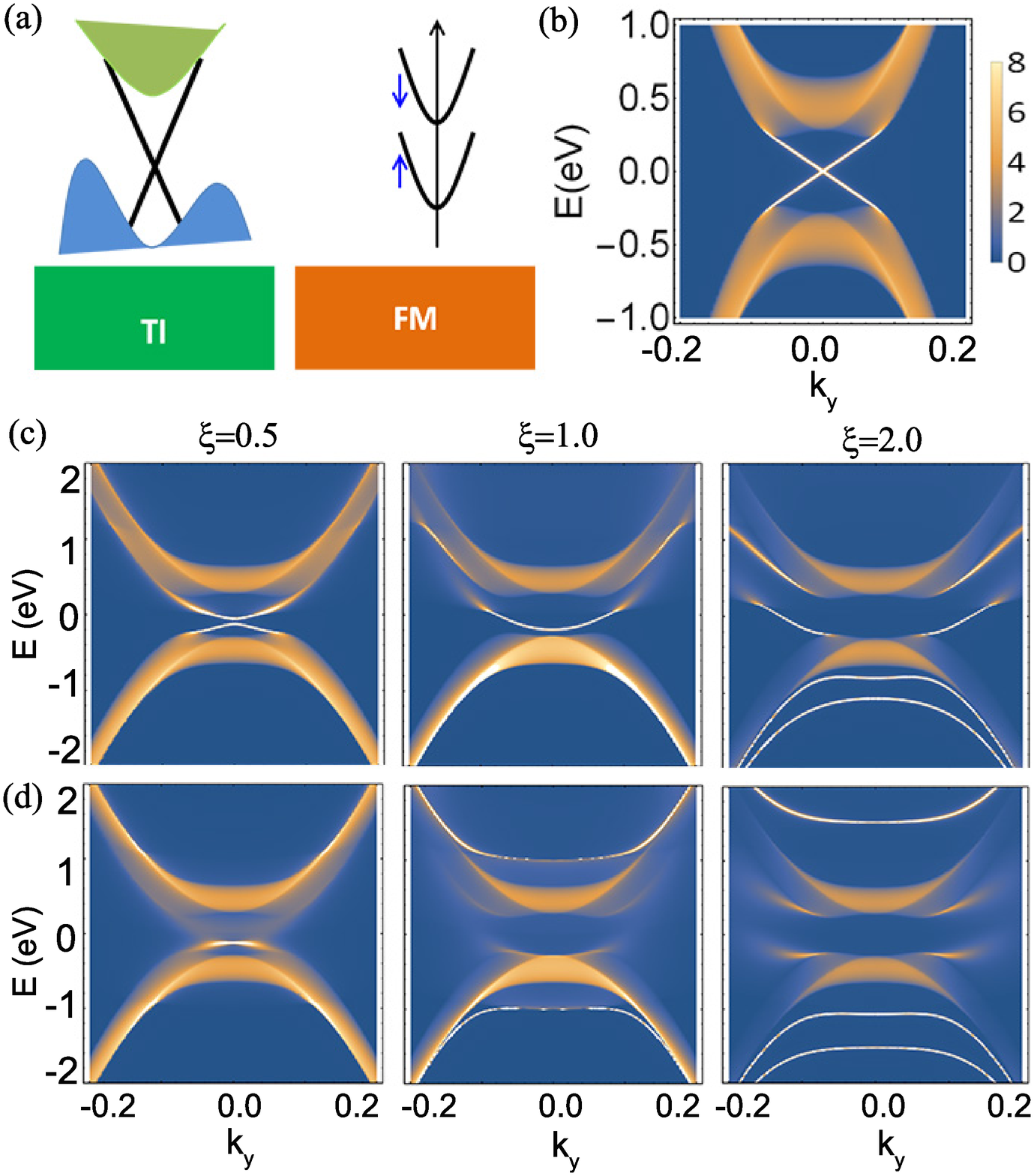}
\caption{ (a) Schematic diagram of the band structure of a topological insulator (TI) and a ferromagnet (FM). (b) Bloch spectral density of a stand-alone TI surface. (c) Spectral density of TI/FM insulator interface for $\varepsilon_{0}=3.5$, $t_{0} = -0.5$ eV, $\Delta = 0.5$ eV, and interface coupling strengh $\xi=t/t_0 = 0.5, 1$ and $2$. (d) Spectral density of TI/FM metal interface for $\varepsilon_{0}=2.5$, $t_{0} = -0.5$ eV, $\Delta = 0.5$ eV.} \label{fig4}
\end{figure}

\noindent
\emph{Tight-binding model}:
The band structure obtained from first-principles calculations  give little room to explore different scenarios, e.g., band alignment. For that purpose, we construct a representative tight-binding (TB) model to carry out calculations of general TI/FM interfaces (Fig.~\ref{fig4}(a)). The TI is described using a four band low-energy effective Hamiltonian of Bi$_2$Se$_3$ as derived in ref.34. The basis consists of the $p_z$ states on Bi and Se with total moment $J_{z} = \pm\frac{1}{2}$, lebelled $|P1_{-}^{+}, \frac{1}{2}\rangle$, $|P2_{+}^{-}, \frac{1}{2}\rangle$, $|P1_{-}^{+}, -\frac{1}{2}\rangle$ and $|P2_{+}^{-}, -\frac{1}{2}\rangle$. In this basis, the TI Hamiltonian is
 \begin{equation}
 	H_{TI}  =  \sum_n [H_{n,n} C^{+}_{n} C_{n} + (H_{n,n+1} C^{+}_{n} C_{n+1} + h.c.)]
 \end{equation}
 where $n$ labels layers in the $z$ direction and
 \begin{eqnarray}
 	H_{n,n}({\bf k}) & =&  (M_{2}(k^2_{x}+k^2_{y})+M_{0}+\frac{M_{1}}{2a^2})\Gamma{_5} \\ \nonumber
 	& +&  A_{0}(k_{y}\Gamma{_1}-k_{x}\Gamma{_2}) \\ \nonumber
 	H_{n,n+1}({\bf k}) & = & (-\frac{M_{1}}{4a^2}\Gamma{_5}-\frac{B_{0}}{2a}i\Gamma{_4}).
 \end{eqnarray}
Here $\Gamma_{1,2,3} = \sigma_{1,2,3} \bigotimes \tau_{1}$ and $\Gamma_{4,5} = I \bigotimes \tau_{2,3}$ where $\sigma_{1,2,3}$ and $\tau_{1,2,3}$ are the Pauli matrices in spin and orbital space respectively. The parameters $M_{0,1,2}$, $B_0$ and $A_0$ are taken from ref.34 and listed in Table \ref{tab3}. The lattice constant is assumed to be $a=3.5$ {\AA}.  For simplicity, we ignore the electron-hole asymmetry term in the Hamiltonian.

The ferromagnet is described by a single-orbital TB Hamiltonian on a cubic lattice
$H^{\uparrow,\downarrow}_{FM}({\bf k}) = \varepsilon_{0} \mp \Delta + 2t_{0}(\cos k_{x}a + \cos k_{y}a + \cos k_{z}a)$, where $\varepsilon_0$ is the onside energy, $\Delta$ is the exchange splitting, and  $t_0$ is the hoping integral. The Green's function of TI interfaced with a ferromagnet is evaluated from the Dyson equation, $G_{TI} = G_{TI}^0 + G_{TI}^0\Sigma G_{TI}$, where $G_{TI}^0$ is the Green's function of the stand-alone TI. The self-energy due to the coupling across the interface, $\Sigma=\tau^\dag G_{FM}^0 \tau$,  is determined by the surface Green's function of the ferromagnet $G_{FM}^0$ and the interface coupling matrix $\tau$ which is determined by the interface hopping parameter $t$\cite{Velev:2004}. The band structure is evaluated in terms of the Bloch spectral density, $n(E,{\bf k}) = -\frac{1}{\pi} {\rm Im}[Tr(G_{TI})]$, in the 2D Brillouin zone.

We note that using a two-band effective Hamiltonian for topologically protected states can only be used in a weak coupling regime ($ \xi=t/t_{0}\ll1$). In this case we obtain the well-known expression $H_{eff}=H_{0}+J_{eff}\sigma_z$ with $J_{eff}=\xi^2\Delta$ (see Supplemental Material). This effective Hamiltonian has been widely used to describe TI/FM insulator interfaces\cite{Luo:2013}. In a more general case the four-band model needs to be used to properly describe TI properties.

\begin{table}
\caption{Four-band-model tight-binding parameters for Bi$_2$Se$_3$.} \label{tab3}
\centering
\begin{ruledtabular}
\begin{tabular}{c|c|c|c|c}
  M$_0$(eV)&M$_1$(eV {\AA}$^2$)&M$_2$(eV {\AA}$^2$)&A$_0$(eV{\AA})&B$_0$(eV{\AA})\\
  \hline
  -0.28&6.86&44.5&3.33&2.26
\end{tabular}
\end{ruledtabular}
\end{table}

Fig.~\ref{fig4}b shows the band structure of the stand-alone TI surface indicating that the linear dispersion of the surface state is well reproduced \cite{Liu:2010}. Then we calculate the band structure of the TI/FM interface as a function of the interface coupling strength ($ \xi$). We distinguish two cases by varying the on-site energy (the Fermi energy) of the FM: (i) FM bands lie above the bulk band gap of Bi$_2$Se$_3$; (ii) FM bands lie within the band gap of Bi$_2$Se$_3$. In the first case, the ferromagnet represents a FM insulator. The corresponding TI/FM insulator spectral density is plotted in Fig.~\ref{fig4}c. It is seen that, in the weak coupling region, a band gap opens in the surface states as expected from the two-band model \cite{Luo:2013}. The stronger coupling pushes the surface bands down in energy eventually causing the lower branch of the Dirac cone to merge with the valence band. In the second case, the ferromagnet represents a FM metal with its bands aligned with the surface state (Fig.~\ref{fig4}d). Even a small amount of hybridization smears the surface state and, for moderate to strong hybridization, the surface state is essentially absorbed in the metal states. These results corroborate the first-principles findings.

Overall, our results indicate that the helical surface state of the TI is not robust upon the perturbation introduced by a ferromagnetic metal interface. Our first-principles and model studies of the Bi$_2$Se$_3$/Ni(111) and Bi$_2$Se$_3$/Co(111) interfaces demonstrate that different work functions of the TI and the FM metals lead to a charge transfer across the interface which shifts the topological surface states down about 0.5 eV below the Fermi energy. The hybridization of these surface states with the metal states at the TI/FM metal interface smears out the topological surface states and destroys their helical spin structure (to larger extent in case of Ni as compared to Co). The band alignment of Bi$_2$Se$_3$ and Ni (Co) places the Fermi energy far in the conduction band of the bulk Bi$_2$Se$_3$, where there are no Dirac states and where the hybridization between the TI and FM states aligns the electron’s spin with the ferromagnet’s magnetization. Our results indicate that the physical picture characterizing the coupling between the TI and the FM metals and its effect on the topological surface states is more intricate than the common notion of a magnetic exchange field breaking the time reversal symmetry and opening the gap in the surface states. These considerations have to be taken into account when interpreting the experimental data and designing applications contingent on the helical spin structure of the TI surface states.

\emph{Acknowledgments}
This work was supported by the National Science Foundation (NSF) through Experimental Program to Stimulate Competitive Research (EPSCoR, Grant No. EPS-1010094). The research at University of Nebraska-Lincoln (UNL) was partly supported by the Nanoelectronics Research Corporation (NERC), a wholly-owned subsidiary of the Semiconductor Research Corporation (SRC), through the Nanoelectronics Research Initiative (NRI) Center for Nanoferroic Devices (CNFD). The research at University of Puerto-Rico (UPR) was partly supported by NSF Grant. No. DMR-1105474. Computations were performed at the UNL Holland Computing Center.

\end{document}